\pdfoutput=1
\documentclass[12pt,conference, onecolumn]{IEEEtran}

\usepackage[numbers]{natbib}
\usepackage{graphicx}
\usepackage{hyperref}
\usepackage{booktabs}
\usepackage{bm}
\usepackage{amsmath}
\usepackage{listings}
\usepackage{xcolor}
\usepackage{subcaption}

\begin{document}

\title{Neural Code Search Revisited: Enhancing Code Snippet Retrieval through Natural Language Intent} 

\author{\IEEEauthorblockN{Geert Heyman}
\IEEEauthorblockA{Nokia Bell Labs\\
Antwerp, Belgium\\
geert.heyman@nokia-bell-labs.com}
\and
\IEEEauthorblockN{Tom Van Cutsem }
\IEEEauthorblockA{Nokia Bell Labs\\
Antwerp, Belgium\\
tom.van\_cutsem@nokia-bell-labs.com}
}

\maketitle

\begin{abstract}
In this work, we propose and study annotated code search: the retrieval of code snippets paired with brief descriptions of their intent using natural language queries. On three benchmark datasets, we investigate how code retrieval systems can be improved by leveraging descriptions to better capture the intents of code snippets. Building on recent progress in transfer learning and natural language processing, we create a domain-specific retrieval model for code annotated with a natural language description. We find that our model yields significantly more relevant search results (with absolute gains up to 20.6\% in mean reciprocal rank) compared to state-of-the-art code retrieval methods that do not use descriptions but attempt to compute the intent of snippets solely from unannotated code.
\end{abstract}

\definecolor{codegreen}{rgb}{0,0.6,0}
\definecolor{codegray}{rgb}{0.5,0.5,0.5}
\definecolor{codepurple}{rgb}{0.58,0,0.82}
\definecolor{backcolour}{rgb}{0.95,0.95,0.92}

\newcommand\tom[1]{\textcolor{red}{\textbf{Tom:~}#1~}}
\newcommand\repourl{\url{https://github.com/nokia/codesearch}}
\newcommand\tagseturl{\url{https://github.com/nokia/codesearch/tree/master/nbs/datasets/so-ds-tags.txt}}


\lstdefinestyle{mystyle}{
    commentstyle=\color{codegreen},
    keywordstyle=\color{magenta},
    numberstyle=\tiny\color{codegray},
    stringstyle=\color{codepurple},
    basicstyle=\ttfamily\footnotesize,
    breakatwhitespace=false,         
    breaklines=true,                 
    captionpos=b,                    
    keepspaces=true,                 
    numbersep=5pt,                  
    showspaces=false,                
    showstringspaces=false,
    showtabs=false,                  
    tabsize=2
}
\lstset{style=mystyle}

\section{Introduction}
Many modern-day applications and software systems are built from a variety of open source libraries and components. As a result, developers find themselves working with a growing number of libraries and tools in their day-to-day coding activities. Quickly and efficiently retrieving relevant documentation for these libraries is becoming increasingly important. Concrete code examples that illustrate the usage of a library or specific library feature are particularly helpful to software developers, as witnessed by the rapid growth of online Q\&A platforms such as Stack Overflow.

To address this need, the research community has been investigating systems that retrieve code snippets from natural language queries~\cite{gu2018deep, sachdev2018retrieval, cambronero2019deep, husain2019codesearchnet, yao2019coacor,Zhangyin:20}. All of these works use neural models to compute the similarity between code snippets and natural language queries. While this setup enables code search on large code corpora, understanding the intent of an unannotated code fragment remains challenging even for experienced developers, let alone for automated systems. For instance,  programmers that were asked to assess the relevance of code snippets to a query reported that it would have been helpful to get additional context about the code snippet~\cite{husain2019codesearchnet}. Moreover, the low percentage of keyword matches between the search terms and the code tokens~\cite{sachdev2018retrieval} makes the retrieval of unannotated code snippets particularly challenging.  

In this paper, we take a step back and investigate code search in a more restricted (but, we argue, equally useful) setting where code snippets are accompanied with brief natural language descriptions that capture their intent. 
Code snippets that are meant to be reused by others are frequently accompanied by descriptions. Such annotated snippets are found for instance in coding cheat sheets, interactive notebook environments, official library documentation, technical blog posts, programming textbooks, or on Q\&A websites such as Stack Overflow. While none of these resources use a common, structured format to label code snippets with their intent, simple heuristics can often be used to pair code snippets with meaningful descriptions. We demonstrate that the quality of the search results improves significantly when these descriptions are taken into account by the snippet ranking algorithm. 

The key contributions of this paper are as follows:
\begin{itemize}
\item We propose the \emph{annotated code search} task: the problem of retrieving code snippets annotated with short descriptions, from natural language queries. To evaluate this task, we created three benchmark datasets, each of which is comprised of an annotated code snippet collection and a set of queries linked to one or more relevant snippets in the collection.
\item We investigate different code retrieval models that make use of code descriptions to rank search results. We propose a new method for fine-tuning a recent NLP transfer learning model to predict the semantic similarity between queries and code descriptions. Using this approach we significantly outperform both classic and neural retrieval baselines.
\item We contrast our results with state-of-the-art retrieval models that rank search results based on the source code only. We find that the descriptions are paramount for obtaining relevant search results on all three benchmarks, even though incorporating the snippet source code itself is still beneficial.
\item  All retrieval models and datasets used in this work are released.\footnote{\repourl}
\end{itemize}

\section{Background \& related work}

\subsection{Code search}
There has been a growing interest in retrieving code using natural language queries~\cite{gu2018deep, sachdev2018retrieval, cambronero2019deep, husain2019codesearchnet, yao2019coacor, Zhangyin:20}.  These works use neural networks to project code and queries in the same vector space. During training, the model parameters are optimized to minimize the distance between pairs of code snippets and corresponding natural language descriptions. These descriptions are typically extracted from the method documentation. The rationale is that, at prediction time, a query will be projected to a vector that is similar to the vector representations of related code snippets. As such, code search can be reduced to a k-nearest neighbors problem in the joint query-snippet vector space. \citet{gu2018deep, sachdev2018retrieval, cambronero2019deep, husain2019codesearchnet, yao2019coacor, Zhangyin:20} have all built retrieval models for snippet collections that consist of pure source code. By contrast, in this work we study a different code search setting where every snippet is annotated with a brief natural language description
describing the code's intent.

\citet{yao2019coacor} proposed a model to automatically generate descriptions from the code and utilize reinforcement learning to directly optimize the descriptions for improving code retrieval. They showed that an ensemble of 1) a model that used the generated descriptions and 2) an existing code search model yielded better performance for retrieving SQL code snippets.  \citet{ye2020} improved on this work with a dual learning model that simultaneously optimizes for code generation and code summarization. Both of these works describe an alternative framework to obtain code descriptions but are otherwise orthogonal to our work. 

Recent work explored the application of NLP transfer learning techniques to code~\cite{kanade2019pre, Zhangyin:20}. CodeBERT~\cite{Zhangyin:20}, a bimodal transformer model for code and natural language obtained state-of-the-art results on highly specific tasks (for example, identifying methods/functions given their documentation among a list of ``distractor'' snippets), but performance on such proxy tasks correlates poorly with performance on actual code search~\cite{husain2019codesearchnet}. It therefore remains to be seen how well CodeBERT performs on downstream tasks such as code search. In this work, we study retrieval of code snippets that leverage both the code tokens \emph{and} a short description of the code. For computing the similarity between the query and the code, we compare with existing models such as NCS~\cite{sachdev2018retrieval} and UNIF~\cite{cambronero2019deep}.

\subsection{Transfer learning in natural language processing}
In the past two years, natural language processing has seen substantial improvements across a variety of tasks including question answering, summarization, and information retrieval~\cite{raffel2019t5, akkalyoncu-yilmaz-etal-2019-applying}. Researchers have shown that fine-tuning large models, pre-trained on vast amounts of unlabeled text data, is often far superior to learning models from scratch on labeled, task-specific data only~\cite[inter alia]{Peters:2018,Devlin:2018}. During pre-training, a neural net is optimized to predict words and/or sentences from their context (i.e., the previous or surrounding words/sentences), and as a result, learns to compute word representations that capture semantic and syntactic properties of the words \emph{and} the context in which they appear. During fine-tuning, the last layer of the pre-trained model is replaced with a task-specific layer and all model parameters are fine-tuned on labeled data of the downstream task. 

All recent state-of-the-art systems use the Transformer architecture~\cite{Vaswani2017}, a neural network that processes sequences using multi-head self-attention. This facilitates capturing long-range dependencies and makes the computations more parallelizable than those of recurrent neural networks (RNNs). Due to the increased scale of the models and text corpora, pre-training has become impractical for the average researcher. Fortunately, several institutions have made pre-trained models publicly available: the universal sentence encoder (USE)~\cite{cer2018universal}, GPT and GPT-2~\cite{radford2018improving,radford2019language}, BERT~\cite{Peters:2018}, RoBERTa~\cite{liu:2019-roberta}, T5~\cite{raffel2019t5} are all publicly available. The models differ in a) the pre-training objectives (e.g., BERT predicts words given the surrounding words, while GPT models only use the previous words); b) the text corpora on which they were trained; c) the model size; and d) variations on how the transformer architecture is used (e.g., BERT computes representations of (sub)words, sentences, and pairs of sentences, whereas the universal sentence encoder is aimed at computing sentence representations only).

Our interest in these transfer learning models stems from the need to compute the semantic similarity between search queries and short code descriptions. Because software development is terminology-heavy, the domains of the training corpora play a crucial role in performance on downstream tasks such as code search. Although the T5 model yields state-of-the-art results on many benchmarks\footnote{\url{https://gluebenchmark.com/leaderboard}}, it is less suited for tasks in the software domain because its pre-training data was filtered with a heuristic to exclude documents that contain code. By contrast, the text corpora on which the universal sentence encoder was trained included data from Q\&A websites such as Stack Overflow.

\section{Annotated code search}\label{sub:acs_problem}

We cast annotated code search as a retrieval task: given a collection of annotated code snippets $\mathcal{C}=s_1,s_2$, ..., $s_N$ where each annotated snippet $s_i$ consists of a code fragment $c_i$ paired with a brief natural language description $d_i$ and a natural language query $q$, the goal is to generate a ranked list of code snippets $s_{q1}$, ..., $s_{qk}$. In the context of this paper, a code snippet will always be paired with a description, so for convenience, we will use the terms \textit{annotated code snippet} and \textit{code snippet} interchangeably. Prior work has studied the retrieval of code snippets without description, we will refer to this setting as \textit{code-only} code search. 

To benchmark the retrieval performance of a model on a snippet collection $\mathcal{C}$, we require a ground truth dataset that links queries to code snippets from $\mathcal{C}$ that match the query.  Performance can then be measured objectively with standard information retrieval metrics: mean reciprocal rank (MRR) and recall@k (r@k). The reciprocal rank is the inverse of the rank of the first matching snippet for a given query, MRR computes the average of the reciprocal ranks for the queries in the evaluation set. For MRR the top-ranked snippets are most influential, e.g., finding a relevant snippet at rank 1 instead of rank 2 will have a bigger impact on the score than finding a snippet at rank 4 instead of rank 5. Recall@k is defined as the percentage of queries for which at least one matching snippet was found in the top-k results.

To get insight into the difficulty of a benchmark, we will also measure the word overlap between the queries and descriptions of matching snippets. We define word overlap between a query $q$ and a description $d$ as the number of unique words that occur in both $q$ and $d$. Relative overlap is computed as the ratio between the overlap and the number of unique words in $q$. Before computing the overlap, $q$ and $d$ are tokenized and lowercased, and stop words are filtered out.

\section{PACS: benchmarks for annotated code search}\label{sec:benchmark}
Existing code search benchmarks~\cite{gu2018deep, sachdev2018retrieval, cambronero2019deep, husain2019codesearchnet, yao2019coacor} link queries to relevant code snippets but these snippets are not consistently paired with descriptions and hence not directly applicable to this study.\footnote{The documentation strings of function snippets in the collections of \citet{gu2018deep,sachdev2018retrieval,cambronero2019deep,husain2019codesearchnet} might be considered a snippet description, but only a minority of the functions in these datasets was documented. Moreover,  \citet{cambronero2019deep} observed that documentation strings are generally a poor representation of the intent of code snippets.} To evaluate and train annotated code retrieval models, we therefore created three benchmarking datasets, which we collectively refer to as the Python Annotated Code Search benchmark (PACS).\footnote{\repourl}  Table \ref{tab:pacs} gives an overview of the PACS datasets. In the remainder of this section, we provide details on the construction of the benchmarks and explain how we obtain relevant training data.

\subsection{Snippet collections and ground truth}
We compile snippet collections from two existing datasets (CoNaLa and StaQC) and mine a third snippet collection from Stack Overflow posts in the data science domain.

\begin{figure}
\centering
\includegraphics[scale=0.33]{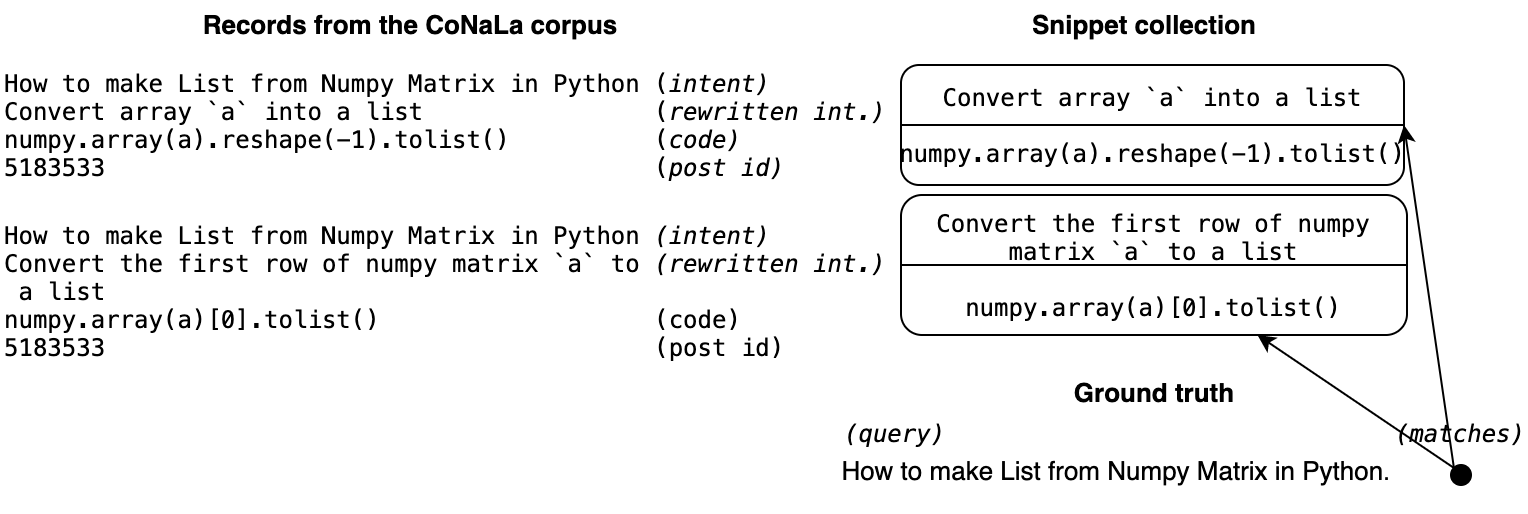}
\caption{Illustration of the CoNaLa corpus~\protect{\cite{yin2018mining}} is converted in a benchmark for annotated code search.}\label{fig:conala}
\end{figure}

\paragraph{CoNaLa}
The CoNaLa corpus \cite{yin2018mining}\footnote{\url{https://conala-corpus.github.io/}} is a curated collection of short Python code snippets annotated with their natural language intent. \citet{yin2018mining} used crowd-sourcing to label a sample of Stack Overflow posts.  An example record of the corpus is shown in Figure \ref{fig:conala}. It consists of a code snippet, intent, re-written intent, and the id of the Stack Overflow post. The code snippets are selected from post answers. The intent is a high-level, natural language description of the code snippet and typically corresponds to the Stack Overflow post title. Because more than one code snippet can be extracted from the same post, different code snippets are sometimes annotated with the same intent. The re-written intent is a reformulation of the intent that should better reflect the full meaning of the code. 

Figure \ref{fig:conala} illustrates how we create a benchmark for annotated code search from the corpus. The snippet collection is constructed by selecting the re-written intents and their corresponding code snippets. This results in 2,777 snippets.  To create the ground truth, we group the CoNaLa records by their Stack Overflow post id. For each group, we use the intent of one of the snippets as a query. The matching snippets are the code snippets that were constructed from the records in this group. Many of the re-written intents are similar to the original intent. Queries that have a relative word overlap of more than 0.5 with the description of a matching snippet are removed. This results in a ground truth dataset with 762 queries.

\begin{figure}
\centering
\includegraphics[scale=0.33]{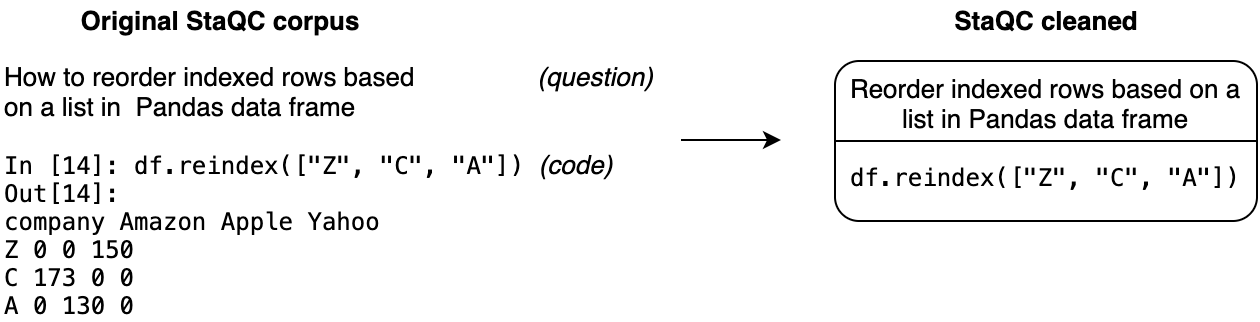}
\caption{Illustration of the additional cleaning for the StaQC corpus~\protect{\cite{yao2018staqc}}: we strip prompts, filter out code snippets that do not parse, and automatically rewrite questions as descriptions with simple regular expressions.}\label{fig:staqc}
\end{figure}

\begin{figure}
\centering
\includegraphics[scale=0.33]{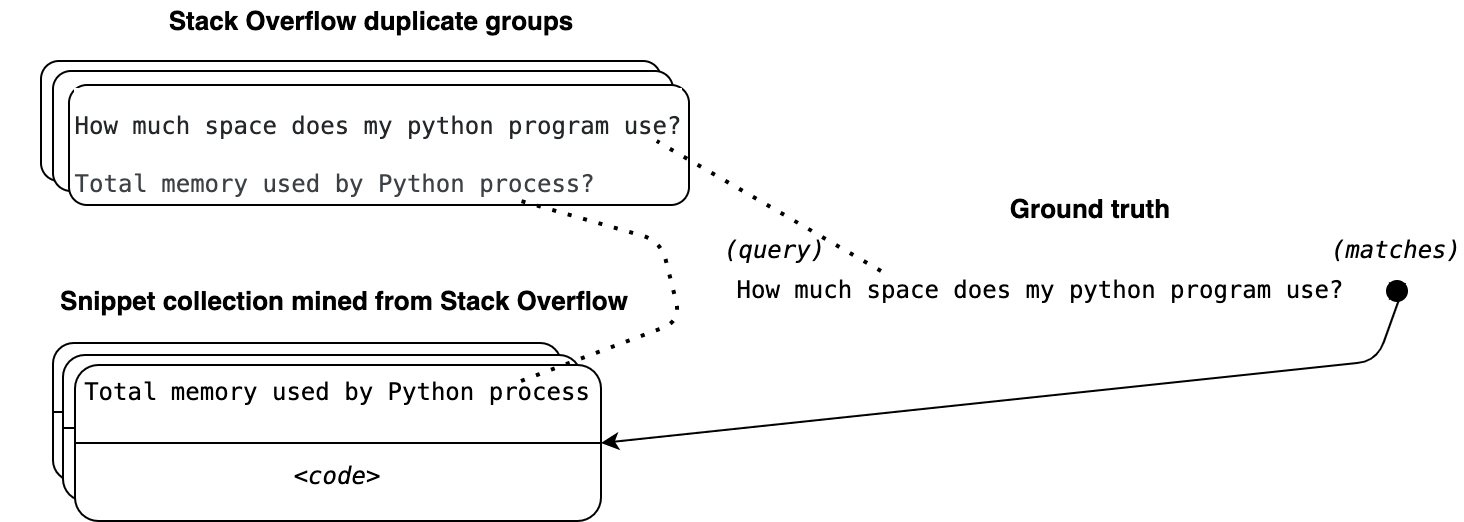}
\caption{To create the StaQC-py and SO-DS ground truth we make use of Stack Overflow posts that were tagged as duplicates by users.}\label{fig:duplicates_gt}
\end{figure}

\paragraph{StaQC-py}
StaQC~\cite{yao2018staqc}\footnote{\url{https://github.com/LittleYUYU/StackOverflow-Question-Code-Dataset}} is a large collection of Python and SQL question-code snippet pairs that are mined from Stack Overflow. These pairs were extracted with machine learning models that identify \textit{how to} questions and their corresponding code snippet(s) in the answers. The dataset has been used in prior work on code summarization \cite{peddamail2018comprehensive} and (code-only) code search \cite{yao2019coacor}. To create a snippet collection we started from the 272K Python question-code pairs in StaQC and performed additional cleaning as illustrated in Figure \ref{fig:staqc}: we strip prompt prefixes ( $>>>$, $\cdot\cdot\cdot$, In [1]: , etc.), filter out code snippets that do not parse in Python 2 and 3, and use simple heuristics to reformulate questions to descriptions. This results in a snippet collection of 204K (description, code snippet) pairs. 

To create the ground truth, we leverage the fact that some Stack Overflow posts are tagged as a duplicate of another post. The titles of duplicate posts are descriptions that are created independently by different users but reflect the same underlying intent. Therefore, as illustrated in Figure \ref{fig:duplicates_gt}, we collect the duplicates of the posts used for the StaQC snippet collection and take the duplicates' titles as queries and the corresponding StaQC snippets as the matching results.

The duplicate posts are collected from the Stack Overflow dump of February 2020.\footnote{Monthly dumps of Stack Overflow can be found here:  \url{https://archive.org/details/stackexchange}} Some posts are tagged as a duplicate of multiple questions, and it also happens that a post A is tagged as a duplicate of post B, while post C is tagged as a duplicate of A. We transitively group all such duplicate relations (i.e., A, B, and C will be part of the same group) and end up with 195K duplicate groups. Most groups consist of two posts, but popular posts can have more than 100 duplicates. To create the StaQC ground truth, we select all groups containing  1) a post that was the source for a StaQC snippet (to have snippets that match the query), as well as 2) a post that was not used for the StaQC data (to select the query). This resulted in 5,497 duplicate groups. From each such group we extracted a query with matching results. The ground truth set is split evenly in a validation and a test set. For the validation set we additionally filter out any queries that overlap with the CoNaLa and SO-DS test sets to avoid overfitting when tuning model hyper-parameters.

\paragraph{SO-DS}
The CoNaLa and StaQC benchmarks have different merits: the CoNaLa snippets and descriptions are curated, but the collection is small and the queries in the ground truth were not created independently from the descriptions. The StaQC benchmark, on the other hand, has a large snippet collection with realistic ground truth queries that were written independently from the snippet descriptions, but despite StaQC's advanced mining methodology the snippet collection contains more noise. We therefore collect a third snippet collection, SO-DS, where we control the quality by only mining from Stack Overflow posts with a large number of upvotes. More specifically, we mine snippets from the most upvoted Stack Overflow posts that are labeled with ``python'' and one or more tags related to data science libraries such as ``tensorflow", ``matplotlib" and ``beautifulsoup".\footnote{We check against a manually curated set of tags that can be consulted here: \tagseturl}~\footnote{We focus on the data science domain because it provides a rich, rapidly evolving open source library ecosystem. Data scientists often need a combination of libraries for data crawling and cleaning, model training and validation, and data visualization needs. As such it is an area of software development where code search tools can add significant value.} 

We iterate over each tag's posts in decreasing order of their upvote scores. For each post, we  extract a maximum of two snippets from the post answers, prioritizing answers with more votes. A snippet is extracted by concatenating all the parseable code blocks from a post answer. We opted to not automatically select the most relevant lines of code with a machine learning model, as was done for StaQC, because incorrect predictions would introduce noise. Note that, unlike CoNaLa and StaQC this dataset is specifically intended for code search, not for code generation or code summarization, where a one-to-one alignment between description and source code is important. Similar to the StaQC benchmark, we use the post title as the snippet description and apply the same cleaning pipeline: we remove prompts, filter out snippets that can not be parsed, and use simple heuristics to reformulate questions to descriptions. When we extracted 250 snippets or ran out of posts with the correct tag, we move on to the next tag.  As a final step, we filter out duplicate snippets, which occur because some Stack Overflow posts with multiple tags were mined more than once. The resulting snippet collection consists of 12,137 snippets and 7,674 unique descriptions.  Table \ref{tab:example_snippets} displays a few examples from the collection. 

The ground truth is collected analogously to StaQC-py by creating queries from Stack Overflow duplicate posts. This results in 2,225 annotated queries, which are evenly split in a validation and test set. The validation set is again filtered to ensure that no queries overlap with the CoNaLa and StaQC-py test sets.

\begin{figure}
\begin{subfigure}{.47\textwidth}
  \centering
  \includegraphics[width=\linewidth]{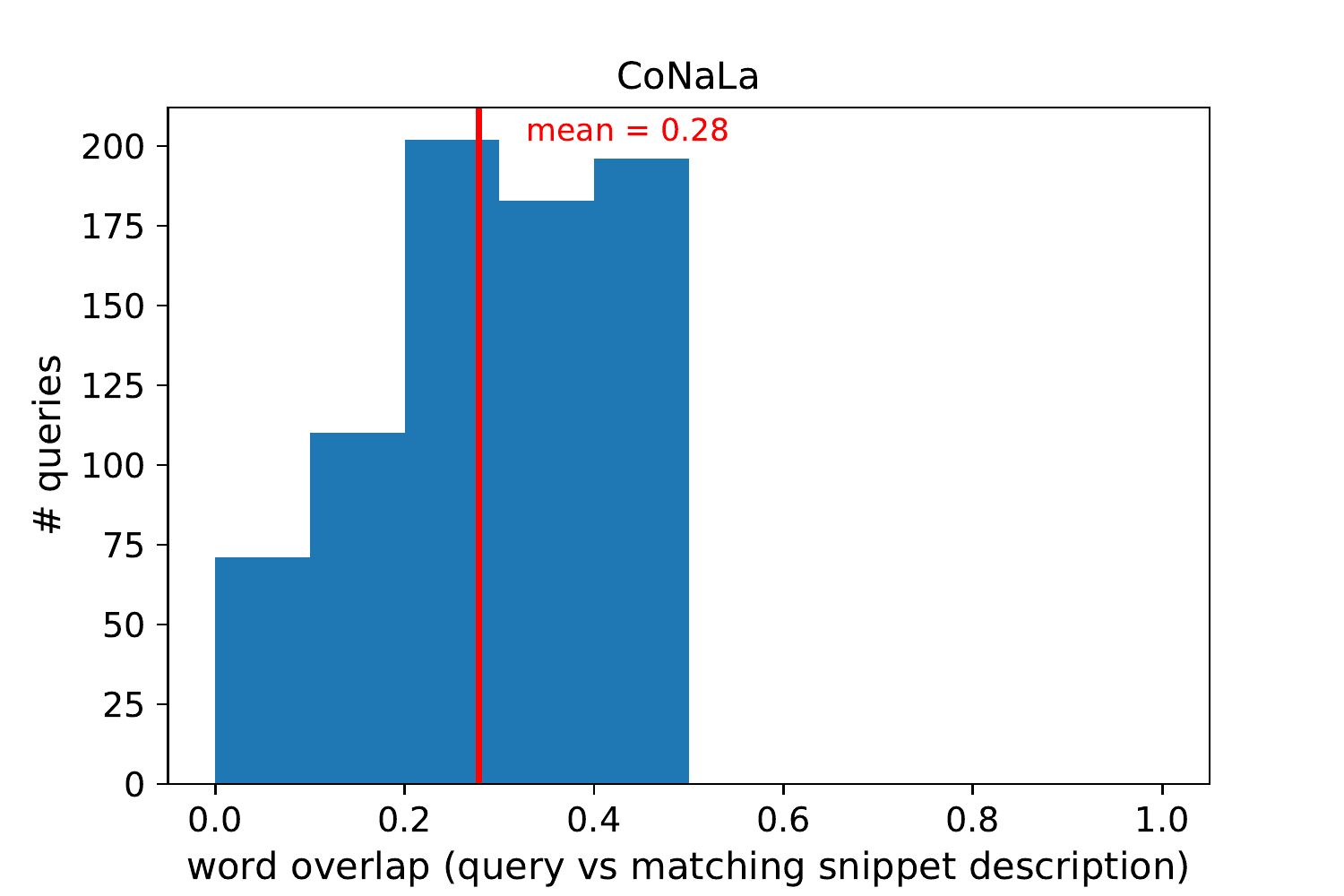}  
  \label{fig:query-overlap-conala}
\end{subfigure}
\begin{subfigure}{.47\textwidth}
  \centering
  \includegraphics[width=\linewidth]{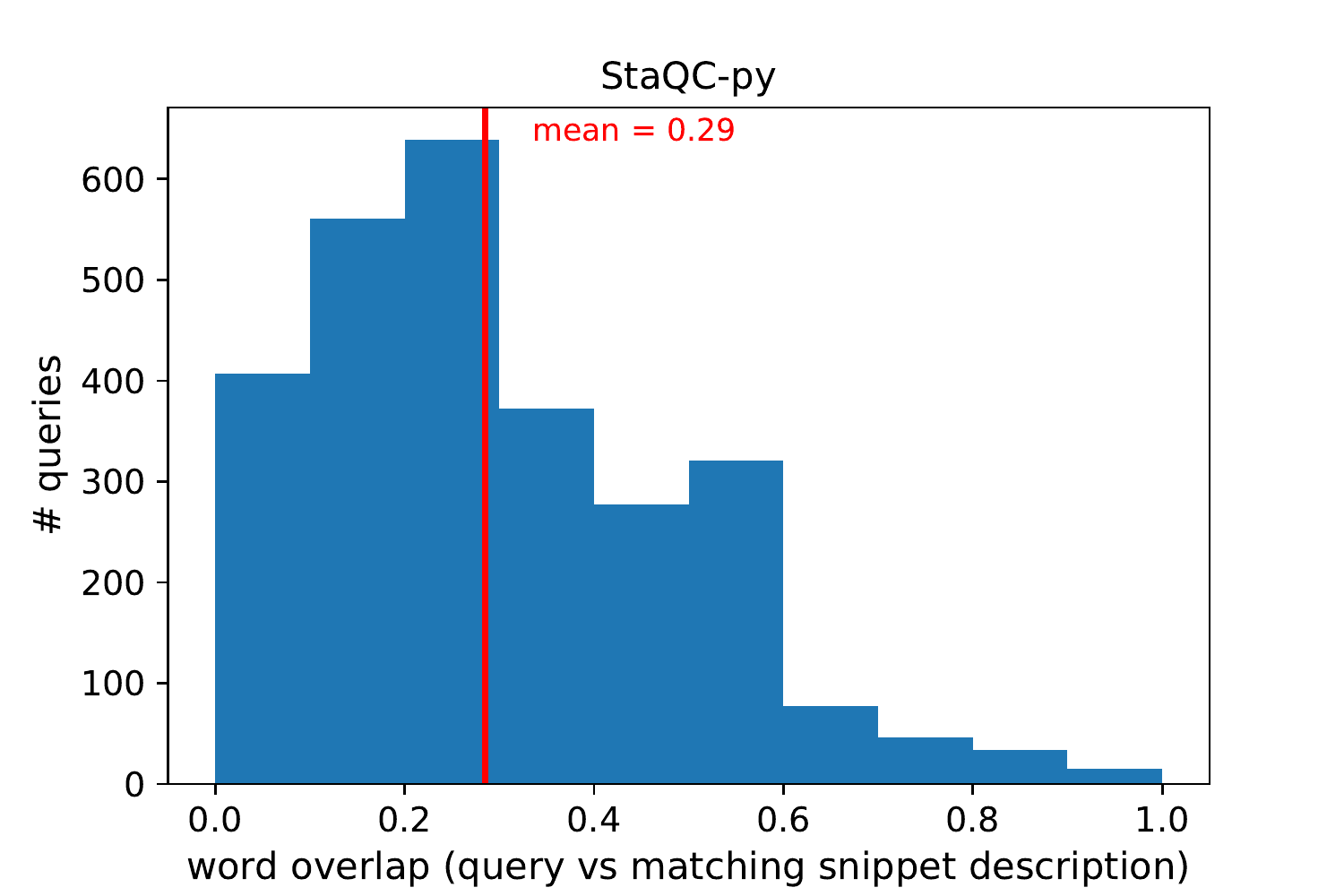}  
  \label{fig:query-overlap-staqc}
\end{subfigure}
\begin{subfigure}{.47\textwidth}
  \centering
  \includegraphics[width=\linewidth]{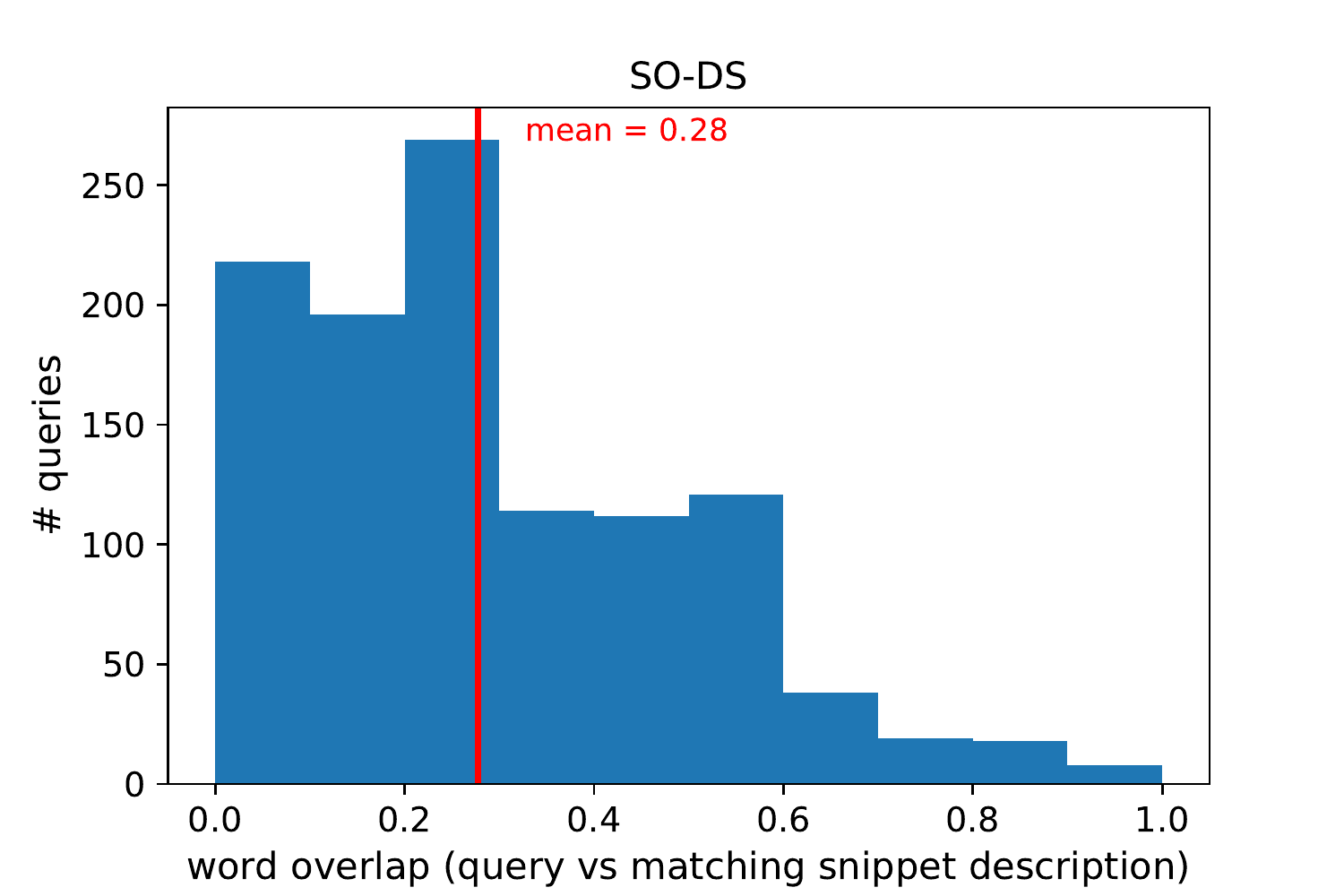}  
  \label{fig:query-overlap-so-ds}
\end{subfigure}
\caption{Histograms of the relative word overlap between the queries and the matching snippet descriptions in the PACS test sets.}\label{fig:query_overlap}
\end{figure}

\paragraph{Benchmark statistics}
Table \ref{tab:pacs} reports summary statistics on the three benchmarks. The CoNaLa snippets are mostly one-liners and its descriptions are on average slightly longer than those in the other collections. The SO-DS snippets are the longest, which is to be expected since the SO-DS mining procedure does not attempt to automatically extract the most relevant lines of code in Stack Overflow posts, as was done for StaQC. Concerning the different ground truth sets, we observe that most of the CoNaLa queries only have one matching snippet, whereas SO-DS and StaQC on average have 1.7 and 3.4 matching snippets. To assess the difficulty of each benchmark, Figure \ref{fig:query_overlap} shows histograms of the relative word overlap between the queries and the descriptions of matching snippets. The three benchmarks have an average relative word overlap between 0.28 and 0.29 and there are queries with a relative overlap of more than 0.6. \footnote{Recall that in order to create the CoNaLa ground truth, we discarded all queries for which the relative overlap with the descriptions of their matching snippets was more than 0.5.}

\begin{table}[]
\centering
\small
\begin{tabular}{@{}llllllll@{}}
\toprule
          & \multicolumn{3}{c}{\textbf{snippet collection}}                                  & \multicolumn{2}{c}{\textbf{test set}}                 \\ \cmidrule(l){2-4} \cmidrule(l){5-6}
\textbf{Benchmark} & \textbf{\#snippets} &\textbf{ \#descr. tokens} & \textbf{code length (\#LoC)} & \textbf{\#queries} & \textbf{\#snippets/query} \\ \midrule
CoNaLa    & 2,777             & 10.32                         & 1.07                & 762              & 1.17                      \\
StaQC-py  & 203,700           & 8.36                          & 9.84                & 2,749            & 3.40                      \\
SO-DS     & 12,137            & 7.35                          & 14.98               & 1,113            & 1.70                      \\ \bottomrule
\end{tabular}
\caption{Overview of the Python annotated code search (PACS) benchmarks.\label{tab:pacs}}
\end{table}

\subsection{Training data}\label{sec:training_data}
Because it is infeasible to construct a code search ground truth dataset that is large enough for model training, we resort to different proxy datasets. We train query-code similarity models (i.e., the code-only models that will be presented in Section \ref{sec:code-sim} ) on the description-code snippet pairs in the respective collections. To train the query-description similarity models (i.e., the description-only models that will be presented in Section \ref{sec:descr_sim}), we create two additional datasets. Firstly, we collect all Python post titles on Stack Overflow before February 2020, resulting in a total of 1.30M questions. Secondly, we create pairs of related code descriptions from the Stack Overflow duplicate post records that were not used to compile ground truth datasets and snippet collections. To not bias questions with many duplicates (e.g., \emph{How to clone or copy a list} has more than 289 duplicates), we only sample one pair per duplicate group. This resulted in 187K pairs of related code descriptions. Table~\ref{tab:so_duplicates} shows a few examples from the dataset.

\begin{table}
\small

\begin{tabular}{p{0.48\linewidth} | p{0.48\linewidth}}

\textbf{Extracting and saving video frames} & \textbf{Log keras output to a file}\\

\begin{lstlisting}[language=Python]
import cv2
vidcap = cv2.VideoCapture('big_buck_bunny_720p_5mb.mp4')
success,image = vidcap.read()
count = 0
while success:
  # save frame as JPEG file   
  cv2.imwrite("frame%d.jpg" % count, image)       
  success,image = vidcap.read()
  print('Read a new frame: ', success)
  count += 1
\end{lstlisting}
& 
\begin{lstlisting}[language=Python]
from keras.callbacks import CSVLogger

csv_logger = CSVLogger('log.csv', 
	append=True, separator=';')
model.fit(X_train, Y_train, callbacks=[csv_logger])
\end{lstlisting}
\\
\url{https://stackoverflow.com/questions/33311153} &\url{https://stackoverflow.com/questions/38445982} \\
\midrule
&\\
\textbf{Plot correlation matrix using pandas} & \textbf{Regex find all overlapping matches} \\

\begin{lstlisting}[language=Python]
import matplotlib.pyplot as plt

plt.matshow(dataframe.corr())
plt.show()
\end{lstlisting}& 
\begin{lstlisting}[language=Python]
import regex as re
s = "123456789123456789"
matches = re.findall(r'\d{10}', s, overlapped=True)
for match in matches: print(match)
\end{lstlisting}
\\
 \url{https://stackoverflow.com/questions/29432629} & \url{https://stackoverflow.com/questions/5616822} \\
\end{tabular}
\caption{Examples from the SO-DS snippet collection. }\label{tab:example_snippets}
\end{table}

\begin{table}[]
\small
\centering
\begin{tabular}{@{}ll@{}}
\toprule
\normalsize{\textbf{Original}}                                   & \normalsize{ \textbf{Duplicate}}                                                      \\ \midrule

Python functions call by reference                  & Is it possible to make a function that modifies the input               \\
Scrape / eavesdrop AJAX data using Javascript?      & Intercept AJAX responses in Chrome Extension                            \\
How to write conditional import statements in QML?  & QML import later module version only if available                       \\
Boost.Asio SSL thread safety                        & Boost Asio and usage of OpenSSL in multithreaded app                    \\
Is it possible to stop Javascript execution?        & System.exit in javascript  \\                      
\bottomrule                      
\end{tabular}
\caption{Examples of Stack Overflow post titles that were tagged as duplicates.}
\label{tab:so_duplicates}
\end{table}

\section{Models for annotated code search}\label{sec:retrieval_model}
In line with recent work in code search~\cite{gu2018deep, sachdev2018retrieval, cambronero2019deep, husain2019codesearchnet, yao2019coacor}, we formulate snippet retrieval as a k-nearest neighbor problem: queries and snippets are projected in a shared vector space such that any given query and its matching snippets are represented by similar vectors. The snippet collection $s_1$, $s_2$, ..., $s_N$  can then be ranked w.r.t. a query $q$ by sorting snippets according to the cosine similarity between their respective vectors $\mathbf{s_1}$, $\mathbf{s_2}$, ..., $\mathbf{s_N}$ and the query vector $\mathbf{q}$.
 
 $$ cos(\mathbf{s_i}, \mathbf{q}) = \frac{ \mathbf{s_i} \cdot \mathbf{q}}{ ||\mathbf{s_i}|| \: ||\mathbf{q}||} $$

An annotated code snippet $s$ consists of a description $d$ and the actual code snippet $c$. Ideally, we would train a model that jointly embeds $d$ and $c$. However, as mentioned in Section \ref{sec:training_data}, we lack sufficient ground truth data to train such a model. Instead, we independently train two embedding models that separately capture the similarity between 1) descriptions and queries (Section \ref{sec:descr_sim}); and 2) code fragments and queries (Section \ref{sec:code-sim}). In Section \ref{sec:ensemble}, we explain how these models can be efficiently combined in a single ensemble model.


\begin{figure}
\includegraphics[width=\textwidth]{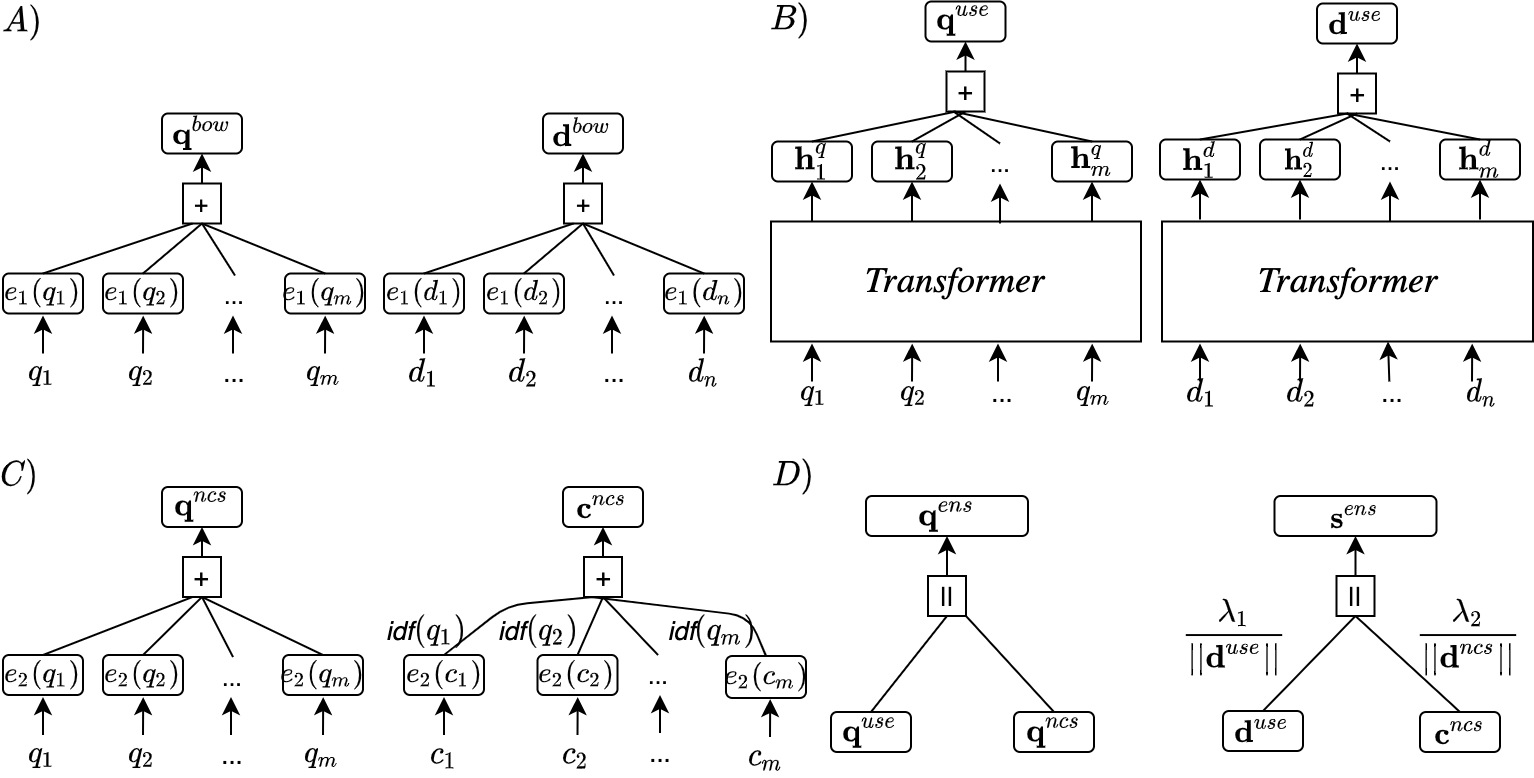}
\caption{Architectures of query/snippet similarity models presented in Section \ref{sec:retrieval_model}: A) depicts the neural bag-of-words model of queries and snippet descriptions; B) depicts the query and snippet description embedding using USE; C) depicts the NCS architecture for embedding queries and actual code; and D) depicts an ensemble of embedding retrieval models.}
\label{fig:codesearchModels}
\end{figure}

\subsection{Query-description similarity}\label{sec:descr_sim}
In this section, we present two query-description similarity models: Neural bag-of-words, which will be used as a baseline, and the universal sentence encoder. 

\subsubsection*{Neural bag-of-words (NBOW)}
Bag-of-words models treat input texts as unordered word sets. While throwing away the sequence order may seem a crude approximation, for information retrieval it results in hard-to-beat baselines. Figure \ref{fig:codesearchModels} A) illustrates how bag-of-words is applied in a neural setting. We obtain embeddings $\mathbf{q^{bow}}$  and $\mathbf{d^{bow}}$ for the query $q$ and description $d$ by summing their respective word embeddings: 

$$\mathbf{q^{bow}} = \sum_{j=1}^m e_1(q_j), \mathbf{d^{bow}} = \sum_{j=1}^n e_1(d_j) $$

Where $q_j$ and $d_j$ denote the j$^{th}$ words in the query/description after pre-processing, and $e_1(\cdot)$ is the word embedding model that maps a description/query word to its embedding.

We pre-process queries and descriptions by applying tokenization, lemmatization, and stopword removal. To learn the word embeddings, we used fastText's~\cite{Bojanowski2016} continuous skip-gram model with negative sampling. This model is similar to the skip-gram model from the word2vec toolkit, but in contrast to the latter, fastText also incorporates subword information. The fastText embeddings are trained on the Python Stack Overflow questions dataset (see Section \ref{sec:training_data}).

\subsubsection*{Universal sentence encoder (USE)}
The universal sentence encoder (USE) is a transformer-based transfer learning model that computes a vector for a sequence of words (e.g., a sentence or a paragraph).\footnote{\citet{cer2018universal} also experimented with a simpler architecture that is also referred to as USE. In this work, we always mean the transformer variant when referring to USE, which obtained the highest scores.} We leverage USE to compute representations for the queries and snippet descriptions as shown in Figure \ref{fig:codesearchModels} B): USE utilizes transformer to obtain contextualized representations $\mathbf{h^q_1}, \mathbf{h^q_2}, ..., \mathbf{h^q_m}$ for each query token. An embedding for the query is then calculated by summing the transformer's outputs $\mathbf{h^q_1}, \mathbf{h^q_2}, ..., \mathbf{h^q_m}$. It is worth noting that each representation $\mathbf{h^q_i}$ that is produced by transformer is contextualized on all the query tokens, not solely from the current and preceding tokens $q_1$, ..., $q_i$ as would be the case with RNNs. The embeddings of the descriptions are computed analogously, with the same architecture and model parameters.

For our experiments, we used a pre-trained model available on TensorFlow Hub\footnote{\url{https://tfhub.dev/google/universal-sentence-encoder/4}} that outputs 512-dimensional embeddings and has a total of 147M parameters.  The model was pre-trained in a multi-task learning setting with three tasks: 1) skip-thought's next sentence prediction~\cite{kiros2015skip}; 2) a conversational input-response task~\cite{henderson2017efficient}, where to goal is to assess the relevance of a response on a given input (e.g., a reply to an email or a Stack Overflow post); and 3) natural language inference using the SNLI corpus~\cite{snli:emnlp2015}. The training data for tasks 1 and 2 comes from a variety of sources crawled from the web, including Stack Overflow. This makes USE a good choice for downstream tasks in the software development domain, such as code search.

Although USE's pre-training ingests textual data from the software domain, without fine-tuning the model does not significantly outperform our baselines (see Section \ref{sec:results}). Training USE directly on code search is infeasible due to the lack of annotated data. Therefore, we propose the detection of duplicate Stack Overflow posts as a proxy task. That is, we tune the USE model to distinguish similar Stack Overflow post titles from unrelated title pairs. This task stimulates that titles of duplicate posts are mapped to similar vectors, while vectors of random title pairs are adjusted to be more orthogonal. As such, the USE embeddings are tailored to semantic similarity computation in the software domain.

A dataset of duplicate title pairs was created as described in Section \ref{sec:benchmark}. For every ``positive'' instance, we construct five negative examples by sampling random title pairs (for which we verify that they were not tagged as each other's duplicate). Given a title pair $(t_1, t_2)$, we model the probability that $t_1$ and $t_2$ are duplicates by first computing the cosine similarity of their USE embeddings and clipping negative similarities to zero with the rectified linear unit (ReLU) activation function, see Equation \ref{eq:sim}. Next, the result is fed to a \textit{layer} with weight parameter $w$, bias $b$ and the logistic activation function $\sigma$, see Equation \ref{eq:prob}. We fine-tune $w$, $b$, and the parameters of USE $\bm\theta$ to minimize binary cross-entropy. The inclusion of the ReLU activation avoids that the training objective pushes the cosine similarity between embeddings of unrelated titles to $-1$, as this corresponds to having a strong negative correlation between the embeddings. 
\begin{align}
sim(t_1, t_2) = ReLU(\;cos( USE(t_1; \bm{\theta}) , \; USE(t_2; \bm{\theta}) )\;)  \label{eq:sim}\\
P( duplicate=1 |  t_1, t_2) = \sigma ( sim(t_1, t_2)\; w \;  + b) ) \label{eq:prob}
\end{align}

When $w$, $b$ are initialized randomly, we found that the model sometimes degenerates to labeling every pair as unrelated. This issue is avoided by initializing $w$, $b$ such that, initially, all pairs with a high cosine similarity are considered a duplicate with high probability. For our experiments, we initialized $w$ and $b$ with $15$ and $-5$ respectively. 

\subsection{Query-code similarity}\label{sec:code-sim}
To compute the similarity between query and code, we experimented with models whose designs are based on two recently published  (unannotated) code search models: NCS~\cite{sachdev2018retrieval} and UNIF~\cite{cambronero2019deep}.

\subsubsection*{NCS}
The architecture for the neural code search (NCS) model~\cite{sachdev2018retrieval} is shown in Figure \ref{fig:codesearchModels} C). Like the neural bag-of-words model introduced in the previous section, the query representation is computed as a sum of the query word embeddings: $\bm q^{ncs} = \sum_{j=1}^m e_2(q_j)$. Similarly, an embedding for the code is computed by summing the code token embeddings weighted with their respective IDF scores: $\bm c^{ncs} = \sum_{j=1}^n idf(c_j)\;e_2(c_j)$. Where $e_2(\cdot)$ is the embedding model that maps code tokens and query words to their respective embeddings. 

\begin{figure}
\includegraphics[width=\textwidth]{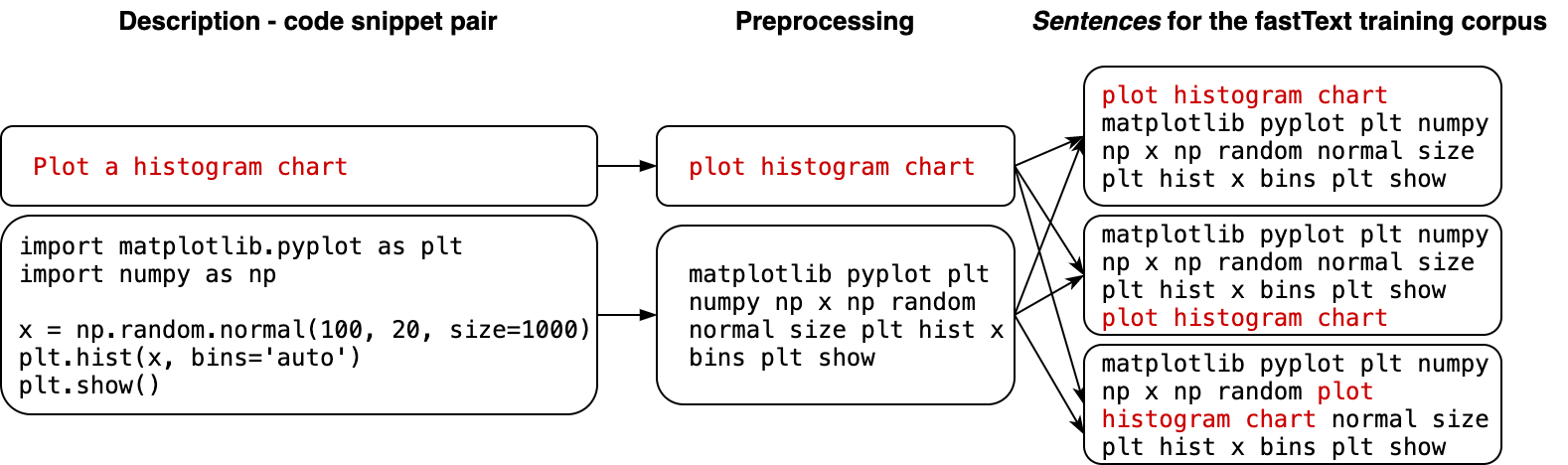}
\caption{Illustration of how \emph{sentences} are created from description-code snippet pairs to serve as input for training fastText embeddings. If we only append the code tokens to the description words (top right) as is done in the original NCS model, \emph{hist} and \emph{histogram} appear far apart. By creating the other contexts (middle and bottom right) they appear in closer proximity, causing fastText to push their embeddings closer in the vector space. }
 \label{fig:ncs_preprocessing}
\end{figure}

The embedding model $e_2(\cdot)$ is trained with fastText  on a multi-modal text corpus containing both source code and natural language code descriptions. For each of the PACS benchmarks, we train a fastText skip-gram model using the respective snippet collections to create the text corpora. In the original NCS model, each \emph{sentence} in the training corpus is constructed from a different description-code snippet pair by concatenating the description words with the code tokens. We found it to be beneficial to further augment the corpora by also: 1) inserting the description words in the middle of the code tokens; and 2) appending the description words to the code tokens (see Figure \ref{fig:ncs_preprocessing}). We also increase fastText's window size parameter to 20 whereas the original model kept fastText's default value (i.e., 5). These modifications are aimed at better capturing the relations between code tokens and description words and are inspired by work on multilingual embedding models~\cite{Vuli2016}. In particular, by enforcing that code tokens and description words that belong to the same snippet appear close to each other in the fastText training corpus, their representations will become more similar.

We pre-process queries by applying tokenization and stopword removal. Code snippets are tokenized with TreeSitter\footnote{\url{https://github.com/tree-sitter/tree-sitter}} and we retain only code identifiers (method names, imports, variable names, keyword argument names) other code identifiers\footnote{The original NCS model did not keep variable and keyword argument names but we found that including these improves results on our benchmarks.}, and inline comments. The code identifiers are split on camel casing and underscores.

\subsubsection*{UNIF}
The UNIF~\cite{cambronero2019deep} and NCS architectures are similar, but instead of using TF-IDF weighting, UNIF computes the code token weights with a neural network.  More specifically, the representation for the code $\bm c^{unif}$ is calculated as a weighted sum of code token embeddings $e_3(c_1)$, $e_3(c_2)$, ... , $e_3(c_m)$:

\begin{align*}
\bm c^{unif} = \sum_i^m w_i \; e_3(c_i)\\
w_i = \frac{exp( e_3(c_i) \cdot \bm a)}{\sum_j^m exp(e_3(c_j) \cdot \bm a)}
\end{align*}

Where $\bm a$ is a vector with the same dimensionality as the embeddings.  UNIF will train $\bm a$ and fine-tune the description and code token embeddings by maximizing the margin between 1) the cosine distance of descriptions and corresponding code snippets; and 2) the cosine distance of descriptions and random code snippets. 

Although making the bag-of-words assumption yields a rudimentary view on source code, state-of-the-art results on code search are obtained under this assumption~\cite{sachdev2018retrieval, cambronero2019deep, husain2019codesearchnet}. We hypothesize that one reason may be that for many snippets in the collections, re-ordering function calls does not lead to a \textit{meaningful} snippet with a different intent (i.e., solving a different task).

\subsection{Ensemble}\label{sec:ensemble}
We expect that the different similarity models we presented will make different types of errors. Therefore,  an ensemble model that computes query-snippet similarity as a linear combination of the fine-tuned USE and NCS model outputs should obtain better results. We note that an ensemble of cosine similarity models can be easily reformulated as a single cosine similarity model:

\begin{align*}
&sim(s, q) = \lambda_1 cos(\bm d^{use}, \bm q^{use}) +  \lambda_2 cos(\bm{c}^{ncs}, \bm q^{ncs}) \\
\text{is equivalent to}\\
&sim(s, q) = cos(\bm s^{ens}, \bm q^{ens})\\
&\bm s^{ens} = <\frac{\lambda_1}{||\bm d^{use}||} \bm d^{use},  \frac{\lambda_2}{||\bm c^{ncs}||}\bm c^{ncs}> \\
&\bm q^{ens} = < \bm q^{use},  \bm q^{ncs} >
\end{align*}

This implementation trick is relevant when scaling up to very large snippets collections, as it allows using approximate nearest neighbor libraries such as faiss~\cite{JDH17}.\footnote{\url{https://github.com/facebookresearch/faiss}} These libraries can do nearest neighbor search with billions of instances but expect that each respective instance/query is represented by a single vector.

\section{Experimental setup}

\subsection{Baselines}
To compare with classic information retrieval approaches, we set up two Okapi BM25 retrieval models using the rank-bm25 library.\footnote{\url{https://pypi.org/project/rank-bm25/}} Okapi BM25 is a popular bag-of-words ranking function based on exact keyword matching. BM25$_\text{code}$ indexes the code, and serves as a baseline for the neural code models. It uses the same pre-processing as NCS and UNIF, except that we replace each query and code token by its lemma. BM25$_{\text{descr}}$ indexes the code descriptions, and serves as a baseline for the neural bag-of-words and universal sentence encoder models. We also report the results of the original NCS model (i.e., without our proposed adaptations) which we reproduced from \cite{sachdev2018retrieval} (to our knowledge, the model is not publicly available).

\subsection{Hyper-parameters}
For training the embeddings of the neural bag-of-words model, we kept fastText's default parameters except for the number of epochs, which we increased to 300. The USE fine-tuning was implemented in TensorFlow~\cite{tensorflow2015-whitepaper}  using stochastic gradient descent (SGD) with mini-batches of size 512, a learning rate of \texttt{1e-4}, and Adam optimization~\cite{Kingma2014}. We fine-tuned USE for 15 epochs, while tracking the retrieval performance on the SO-DS validation set every $51,200$ examples, after which we retained the best performing snapshot to evaluate on the test set. The NCS embeddings (our version) were trained for 30 epochs, with window size 20, and without filtering out infrequent words. Other parameters were kept at fastText's default setting. The original NCS model did not change fastText's defaults. The UNIF model was trained for 20 epochs, using SGD with mini-batches of size 32, a learning rate of \texttt{1e-4}, and Adam optimization~\cite{Kingma2014}. For StaQC-py and SO-DS, we track retrieval performance on the respective validation sets every epoch and keep the best performing checkpoint for evaluation on the test set.  For CoNaLa, we do not have a validation set so we keep the model after 20 epochs. The $\lambda$ weights for the ensemble model were tuned on the SO-DS validation set and were set to  $0.5$ for NCS and $1$ for USE.

\subsection{Evaluation metrics}
As discussed in Section~\ref{sub:acs_problem}, we use mean reciprocal rank (MRR) and recall@k (r@k) to measure snippet retrieval performance. We cut-off the search result list at 10 snippets when computing MRR as users will rarely look beyond the first 10 results. For recall@k, we set $k=3$ and $k=10$, measuring the percentage of queries for which at least one matching snippet was found in the top-3 and top-10 results respectively.

\section{Results \& discussion} \label{sec:results}
\begin{table}[]
\centering
\small
\begin{tabular}{@{}lllllllllll@{}}
\toprule
\textbf{}       & \textbf{}       & \multicolumn{3}{c}{\textbf{CoNaLa}} & \multicolumn{3}{c}{\textbf{StaQC-py}} & \multicolumn{3}{c}{\textbf{SO-DS}} \\ \cmidrule(l){3-5}  \cmidrule(l){6-8}  \cmidrule(l){9-11} 
\textbf{Model}           & \textbf{Inputs} & \textbf{MRR}        & \textbf{r@3 }       & \textbf{r@10 }     & \textbf{MRR}         & \textbf{r@3}       & \textbf{r@10}       & \textbf{MRR}        & \textbf{r@3}       & \textbf{r@10}      \\ \midrule
BM25$_\text{code}$      & code            & 6.9        & 7.0        & 14.6      & 2.0         & 2.1        & 4.2        & 6.7        & 7.7       & 13.5      \\
NCS$_\text{orig}$       & code            & 1.5                              & 1.3                              & 3.7                               & 1.7                              & 1.7                              & 3.5                               & 4.4                              & 4.9                              & 8.9                               \\
NCS$_{\text{ours}}$      & code               & \textbf{16.7}                    & \textbf{19.9}                    & \textbf{31.2}                     & \textbf{3.0}                     & \textbf{3.2}                     & \textbf{6.1}                      & 11.3                    & 12.7                    & 23.5                     \\ 
UNIF            & code            & 15.6       & 18.3       & 29.8      & 2.4        & 2.8        & 5.5        &    \textbf{11.6}        &     \textbf{13.1}      &    \textbf{23.7}       \\ \midrule
BM25$_\text{descr}$     & description     & 23.8       & 26.4       & 39.1      & 7.5         & 8.0        & 13.3       & 21.6       & 23.8      & 32.3      \\
NBOW            & description     & 17.7       & 18.8       & 30.6      & 9.5         & 10.9       & 16.6       & 24.7       & 27.7      & 38.0      \\
USE             & description     & 18.1       & 20.7       & 32.3      & 10.4        & 11.2       & 17.6       & 24.4       & 27.3      & 37.9      \\
USE$_\text{tuned}$      & description     & 30.0       & 34.0       & 48.7      & 11.1        & 11.5       & 19.0       & 29.0       & 32.3      & 45.3      \\
USE$_\text{tuned+ReLU}$ & description     & \textbf{34.0}       & \textbf{37.9}       & \textbf{54.9}      & \textbf{11.7}        & \textbf{12.8} & \textbf{20.3}       & \textbf{30.4}       & \textbf{33.3}      & \textbf{48.5}    \\ \midrule
Ensemble        & code + descr.   & \textbf{35.1}         &    \textbf{39.8}     &    \textbf{57.2}       &     \textbf{12.6}        &      \textbf{14.3}     &      \textbf{22.7}      &           \textbf{32.3} &     \textbf{37.0}      &     \textbf{50.1}      \\ \bottomrule
\end{tabular}
\caption{Results on the Python annotated code search (PACS) benchmark test sets. Mean reciprocal rank (MRR), recall@3 (r@3) and recall@10 (r@10) are expressed as percentages. NCS$_{\text{original*}}$ and UNIF$_*$ are the results obtained with our implementations of the NCS/UNIF models as described in \protect{\cite{sachdev2018retrieval}} and \protect{\cite{cambronero2019deep}}, respectively (the original models were not released). }
\label{tab:results}
\end{table}

The performance of the various models on our annotated code search benchmarks is listed in Table \ref{tab:results}. The results are grouped by the information that the models index: the source code,  the description, or both. As expected, we find that leveraging code descriptions significantly boosts retrieval performance. The big gap between the best code-only and best description-only models confirms that retrieving snippets on the source code alone remains a very challenging task. This is likely due to the fact that the vocabulary of code tokens and description tokens is quite disjoint~\cite{sachdev2018retrieval}. We also observe that our NCS variant and UNIF outperform classic keyword matching. The modifications to NCS that we proposed in Section~\ref{sec:code-sim} improve performance over the original model across all three benchmarks, and achieve results on par with or better than UNIF. Table \ref{tab:ncs_ablation} reports an ablation study of the proposed adjustments. It demonstrates that fastText hyper-parameter tuning, the more inclusive code pre-processing (i.e., retaining all code identifiers), and the fastText training corpus augmentation illustrated in Figure \ref{fig:ncs_preprocessing}, all contribute to this performance increase.  

In an additional experiment, we verified whether using more training data for NCS/UNIF could help close the gap with the description-only models. To this end, we constructed a superset of the SO-DS  snippet collection by mining annotated snippets with the procedure described in Section \ref{sec:benchmark}, except that we retain 1,000 instead of 250 snippets per tag. Somewhat surprisingly, we found that having four times more training data did not improve the performance for NCS and UNIF.

\begin{table}[]
\small
\centering
\begin{tabular}{@{}lccccccccc@{}}
\toprule
                                        & \multicolumn{3}{c}{\textbf{CoNaLa}}                                                                     & \multicolumn{3}{c}{\textbf{StaQC-py}}                                                                   & \multicolumn{3}{c}{\textbf{SO-DS}}                                                                      \\ \cmidrule(l){2-4}  \cmidrule(l){5-7}  \cmidrule(l){8-10} 
\textbf{Method}                         & \multicolumn{1}{l}{\textbf{MRR}} & \multicolumn{1}{l}{\textbf{r@3}} & \multicolumn{1}{l}{\textbf{r@10}} & \multicolumn{1}{l}{\textbf{MRR}} & \multicolumn{1}{l}{\textbf{r@3}} & \multicolumn{1}{l}{\textbf{r@10}} & \multicolumn{1}{l}{\textbf{MRR}} & \multicolumn{1}{l}{\textbf{r@3}} & \multicolumn{1}{l}{\textbf{r@10}} \\ \midrule
NCS$_{\text{original*}}$                & 1.5                              & 1.3                              & 3.7                               & 1.7                              & 1.7                              & 3.5                               & 4.4                              & 4.9                              & 8.9                               \\
NCS$_{\text{original*+epochs=30}}$      & 4.7                              & 5.0                              & 11.7                              & 2.0                              & 2.0                              & 4.1                               & 6.3                              & 7.3                              & 14.5                              \\
NCS$_{\text{original*+allIdentifiers}}$      & 2.6                              & 2.8                              & 6.0                               & 2.0                              & 2.0                              & 4.3                               & 6.5                              & 7.2                              & 14.4                              \\
NCS$_{\text{original*+window=20}}$      & 1.5                              & 1.7                              & 3.7                               & 2.0                              & 2.1                              & 4.1                               & 5.9                              & 6.0                              & 12.6                              \\
NCS$_{\text{original*+contextAugment}}$ & 2.3                              & 2.6                              & 5.5                               & 1.6                              & 1.5                              & 3.5                               & 7.0                              & 7.0                              & 14.6                              \\ \midrule
NCS$_{\text{ours-epochs=30}}$           & 7.4                              & 8.1                              & 18.1                              & 2.4                              & 2.5                              & 5.3                               & 9.8                              & 11.3                             & 19.2                              \\
NCS$_{\text{ours-allIdentifiers}}$           & 10.1                             & 10.6                             & 19.9                              & 2.5                              & 2.7                              & 5.3                               & 8.9                              & 8.7                              & 19.3                              \\
NCS$_{\text{ours-window=20}}$           & 15.8                             & 18.0                             & 29.0                              & 2.5                              & 2.7                              & 5.4                               & 9.8                              & 11.1                             & 20.7                              \\
NCS$_{\text{ours-contextAugment}}$      & 14.3                             & 15.6                             & 28.5                              & 2.6                              & 2.8                              & 5.3                               & 10.0                             & 11.3                             & 22.0                              \\
NCS$_{\text{ours}}$                     & \textbf{16.7}                    & \textbf{19.9}                    & \textbf{31.2}                     & \textbf{3.0}                     & \textbf{3.2}                     & \textbf{6.1}                      & \textbf{11.3}                    & \textbf{12.7}                    & \textbf{23.5}                     \\ \bottomrule
\end{tabular}
\caption{Ablation study of the changes to the original NCS model: NCS$_{\text{original*+x}}$ corresponds to the original model with one of our modifications, and NCS$_{\text{ours-x}}$ corresponds to our NCS variant without one of our modifications. \textit{contextAugment} refers to the method we propose for generating the fastText input data (see Figure \ref{fig:ncs_preprocessing}).}
\label{tab:ncs_ablation}
\end{table}

When comparing the description-only baselines, we find that on CoNaLA classic keyword matching (i.e., BM25) outperforms the neural baselines (i.e., neural bag-of-words model and USE without fine-tuning), whereas on StaQC-py and SO-DS the inverse is true. This indicates that for StaQC-py and SO-DS there tends to be more (accidental) keyword overlap between queries and descriptions of irrelevant snippets. This is to be expected given that the StaQC-py and SO-DS snippet collections are considerably larger than the CoNaLa collection.

Another important finding is that, across all three benchmarks, the proposed fine-tuning recipe boosts USE's performance significantly.  The fine-tuned USE models outperform all other models with large margins. The experiments also validate the benefit of introducing the ReLU activation before the output layer as it consistently leads to the best results.  Given these large improvements, we also considered fine-tuning other pre-trained transformer models, We conducted preliminary experiments with sentence-BERT models~\cite{reimers-gurevych-2019-sentence}, which outperformed USE on semantic similarity benchmarks, and GPT-2~\cite{radford2019language}.  Even after fine-tuning these models were unable to beat the baselines. For the sentence-BERT models, we attribute this to the mismatch between the domain of the pre-training corpora and the software development domain. For the GPT-2 model, on the other hand, we found that the cosine similarities between hidden states of unrelated sequences are particularly high, a phenomenon that is studied in \citet{ethayarajh2019contextual}. This makes the model a poor fit for our fine-tuning procedure.

A final observation from Table \ref{tab:results} is that the ensemble model, which combines USE$_\text{tuned+ReLU}$ with NCS$_\text{ours}$,  yields the best results. This indicates that the incorporation of the source code is still beneficial. To demonstrate that the ensemble model can answer non-trivial queries,  Table \ref{tab:example_queries} reports a top-3 result from the SO-DS collection for four non-trivial example queries. Although all queries have at most one word in common with the description or code, the results capture the query intent very well. Interestingly, the result for the query \textit{check that tf uses gpu} also uncovers a limitation of current code search systems. The code snippet is relevant but outdated as it will not work in the latest versions of the TensorFlow library. Dealing with the evolution of software libraries and tools in a code search engine is an interesting challenge for future research.

To further improve code search, we see the most potential in learning better code representations. An open question is whether the transfer learning techniques that have pushed the state of the art in natural language processing, can yield similar improvements when leveraged for code representation learning. One could question to what extent the distributional hypothesis (i.e., the idea that words that occur in the same contexts tend to have similar meanings~\cite{Harris:1954}) is as powerful for source code. Because software developers are trained to avoid code duplication (by refactoring repeated code fragments into functions, methods, modules, etc.), what can be learned from token co-occurrence is inherently more limited, due to the sparsity of patterns. This raises the question of whether we should not incorporate other modalities, such as runtime information, when learning code representations.

\begin{table}
\footnotesize

\begin{tabular}{p{0.48\linewidth} | p{0.48\linewidth}}

\textbf{Q:} export plot to png & \textbf{Q:} regex to find urls\\

\textbf{D:} Save plot to image file instead of displaying it using matplotlib & \textbf{D:} Regular expression to extract url from an html link \\

\textbf{C:}
\begin{lstlisting}[language=Python]
import matplotlib.pyplot as plt

# create figure & 1 axis
fig, ax = plt.subplots(nrows=1, ncols=1)  
ax.plot([0, 1, 2], [10, 20, 3])
# save the figure to file
fig.savefig('path/to/save/image/to.png')   
plt.close(fig)    # close the figure
\end{lstlisting}
& 
\textbf{C:}
\begin{lstlisting}[language=Python]
import re

url = '<a href="http://www.ptop.se" target="_blank">http://www.ptop.se</a>'
r = re.compile('(?<=href=").*?(?=")')
r.findall(url)
\end{lstlisting}
\\
\textbf{S:} \url{https://stackoverflow.com/questions/9622163} &\textbf{S:} \url{https://stackoverflow.com/questions/499345} \\
\midrule
&\\
\textbf{Q:} create df from json lines & \textbf{Q:} check that tf uses gpu\\
\textbf{D:} Loading a file with more than one line of json into pandas & \textbf{D:} Tell if tensorflow is using gpu acceleration from inside python shell  \\
\textbf{C:}
\begin{lstlisting}[language=Python]
import pandas as pd

data = pd.read_json(
	'/path/to/file.json', lines=True)
\end{lstlisting} &\textbf{C:} \begin{lstlisting}[language=Python]
sess = tf.Session(config=tf.ConfigProto(log_device_placement=True))
\end{lstlisting}\\
\textbf{S:} \url{https://stackoverflow.com/questions/30088006} & \textbf{S:} \url{https://stackoverflow.com/questions/38009682} \\
\end{tabular}
\caption{Example queries (\textbf{Q}) with one of the top-3 results (a description \textbf{D}, a code snippet \textbf{C}, and a url to the source \textbf{S}) from the SO-DS collection) returned by the ensemble model. }\label{tab:example_queries}
\end{table}

\subsection*{Threats to validity}
\subsubsection*{Stack Overflow titles as queries}
Similar to previous code search studies~\cite{sachdev2018retrieval, cambronero2019deep, yao2019coacor}, ground truth queries are compiled from Stack Overflow post titles. While the resulting evaluation and test sets are a good reflection of the types of questions that programmers have, Stack Overflow titles can be more verbose than actual code search queries.

\subsubsection*{Training data}
Due to the inherent differences between their architectures, NCS/UNIF and USE can not be trained on the same data. Therefore the results are sensitive to the training data we chose for the respective models. As in this work we were mainly concerned with investigating the benefit of code descriptions, we aimed to optimize the training data for the code-only models. We verified if the performance of these models improved when training them with more in-domain code snippet-description pairs but this was not the case.

\subsubsection*{Interpreting performance}
Our approach to construct the respective ground truth datasets does not yield an exhaustive set of all the relevant snippets for a given query.  As such, the performance metrics in this paper are only meaningful to compare the performance of models relative to each other, not to measure absolute performance. For instance, a recall@10 score of 50\% does not reflect that the model retrieves relevant snippets in the top 10 for (only) half of the queries. 

\subsubsection*{Snippet collection}
The choice of the snippet collection unavoidably impacts the results. For instance, one could argue that the lack of static types makes learning code representations for Python more challenging compared to languages like Java. Given the considerable performance gaps between the code-only and description-only models, across three different snippet collections, we expect our conclusions to remain valid on other collections.

\section{Conclusion}

As the number of software libraries and tools increases, so does the need for effective code search tools to identify relevant example code. Most work on code search considers the problem of retrieving unannotated code snippets based on natural language queries. In this work, we considered the problem of retrieving code snippets annotated with a brief description of their intent, which is usually available for code that is part of tutorials, cheat sheets, textbooks, notebooks, and other forms of code documentation.

We showed that by leveraging code descriptions and by applying recent advances in natural language processing, we can build retrieval models that produce far more relevant search results compared to models that solely use source code. In particular, we described how we can fine-tune the universal sentence encoder for code search, showed how it can be effectively combined with a new variant of the neural code search model, and evaluated the models on three new code search benchmarks. 
On these benchmarks, we significantly outperformed state-of-the-art code search models with absolute gains up to 20.6\%, 23.9\%, and 26.4\% in mean reciprocal rank (MRR), recall@3, and recall@10.

\bibliography{library.bib}

\begin{thebibliography}{31}
\providecommand{\natexlab}[1]{#1}
\providecommand{\url}[1]{#1}
\csname url@samestyle\endcsname
\providecommand{\newblock}{\relax}
\providecommand{\bibinfo}[2]{#2}
\providecommand{\BIBentrySTDinterwordspacing}{\spaceskip=0pt\relax}
\providecommand{\BIBentryALTinterwordstretchfactor}{4}
\providecommand{\BIBentryALTinterwordspacing}{\spaceskip=\fontdimen2\font plus
\BIBentryALTinterwordstretchfactor\fontdimen3\font minus
  \fontdimen4\font\relax}
\providecommand{\BIBforeignlanguage}[2]{{%
\expandafter\ifx\csname l@#1\endcsname\relax
\typeout{** WARNING: IEEEtranN.bst: No hyphenation pattern has been}%
\typeout{** loaded for the language `#1'. Using the pattern for}%
\typeout{** the default language instead.}%
\else
\language=\csname l@#1\endcsname
\fi
#2}}
\providecommand{\BIBdecl}{\relax}
\BIBdecl

\bibitem[Gu et~al.(2018)Gu, Zhang, and Kim]{gu2018deep}
X.~Gu, H.~Zhang, and S.~Kim, ``{Deep code search},'' in \emph{2018 IEEE/ACM
  40th International Conference on Software Engineering (ICSE)}.\hskip 1em plus
  0.5em minus 0.4em\relax IEEE, 2018, pp. 933--944.

\bibitem[Sachdev et~al.(2018)Sachdev, Li, Luan, Kim, Sen, and
  Chandra]{sachdev2018retrieval}
S.~Sachdev, H.~Li, S.~Luan, S.~Kim, K.~Sen, and S.~Chandra, ``{Retrieval on
  source code: a neural code search},'' in \emph{Proceedings of the 2nd ACM
  SIGPLAN International Workshop on Machine Learning and Programming
  Languages}, 2018, pp. 31--41.

\bibitem[Cambronero et~al.(2019)Cambronero, Li, Kim, Sen, and
  Chandra]{cambronero2019deep}
J.~Cambronero, H.~Li, S.~Kim, K.~Sen, and S.~Chandra, ``{When deep learning met
  code search},'' in \emph{Proceedings of the 2019 27th ACM Joint Meeting on
  European Software Engineering Conference and Symposium on the Foundations of
  Software Engineering}, 2019, pp. 964--974.

\bibitem[Husain et~al.(2019)Husain, Wu, Gazit, Allamanis, and
  Brockschmidt]{husain2019codesearchnet}
H.~Husain, H.-H. Wu, T.~Gazit, M.~Allamanis, and M.~Brockschmidt,
  ``{CodeSearchNet Challenge: Evaluating the State of Semantic Code Search},''
  \emph{arXiv preprint arXiv:1909.09436}, 2019.

\bibitem[Yao et~al.(2019)Yao, Peddamail, and Sun]{yao2019coacor}
Z.~Yao, J.~R. Peddamail, and H.~Sun, ``{CoaCor: code annotation for code
  retrieval with reinforcement learning},'' in \emph{The World Wide Web
  Conference}, 2019, pp. 2203--2214.

\bibitem[Feng et~al.(2020)Feng, Guo, Tang, Duan, Feng, Gong, Shou, Qin, Liu,
  Jiang, and Zhou]{Zhangyin:20}
Z.~Feng, D.~Guo, D.~Tang, N.~Duan, X.~Feng, M.~Gong, L.~Shou, B.~Qin, T.~Liu,
  D.~Jiang, and M.~Zhou, ``{CodeBERT: A Pre-Trained Model for Programming and
  Natural Languages},'' \emph{arXiv preprint arXiv:2002.08155}, 2020.

\bibitem[Ye et~al.(2020)Ye, Xie, Zhang, Hu, Wang, and Zhang]{ye2020}
\BIBentryALTinterwordspacing
W.~Ye, R.~Xie, J.~Zhang, T.~Hu, X.~Wang, and S.~Zhang, ``{Leveraging Code
  Generation to Improve Code Retrieval and Summarization via Dual Learning},''
  in \emph{Proceedings of The Web Conference 2020}, ser. WWW '20.\hskip 1em
  plus 0.5em minus 0.4em\relax New York, NY, USA: Association for Computing
  Machinery, 2020, pp. 2309--2319. [Online]. Available:
  \url{https://doi.org/10.1145/3366423.3380295}
\BIBentrySTDinterwordspacing

\bibitem[Kanade et~al.(2019)Kanade, Maniatis, Balakrishnan, and
  Shi]{kanade2019pre}
A.~Kanade, P.~Maniatis, G.~Balakrishnan, and K.~Shi, ``{Pre-trained Contextual
  Embedding of Source Code},'' \emph{arXiv preprint arXiv:2001.00059}, 2019.

\bibitem[Raffel et~al.(2019)Raffel, Shazeer, Roberts, Lee, Narang, Matena,
  Zhou, Li, and Liu]{raffel2019t5}
\BIBentryALTinterwordspacing
C.~Raffel, N.~Shazeer, A.~Roberts, K.~Lee, S.~Narang, M.~Matena, Y.~Zhou,
  W.~Li, and P.~J. Liu, ``{Exploring the limits of transfer learning with a
  unified text-to-text transformer},'' \emph{arXiv preprint arXiv:1910.10683},
  2019. [Online]. Available: \url{https://arxiv.org/pdf/1910.10683.pdf}
\BIBentrySTDinterwordspacing

\bibitem[{Akkalyoncu Yilmaz} et~al.(2019){Akkalyoncu Yilmaz}, Wang, Yang,
  Zhang, and Lin]{akkalyoncu-yilmaz-etal-2019-applying}
\BIBentryALTinterwordspacing
Z.~{Akkalyoncu Yilmaz}, S.~Wang, W.~Yang, H.~Zhang, and J.~Lin, ``{Applying
  BERT to Document Retrieval with Birch},'' in \emph{Proceedings of the 2019
  Conference on Empirical Methods in Natural Language Processing and the 9th
  International Joint Conference on Natural Language Processing (EMNLP-IJCNLP):
  System Demonstrations}.\hskip 1em plus 0.5em minus 0.4em\relax Hong Kong,
  China: Association for Computational Linguistics, nov 2019, pp. 19--24.
  [Online]. Available: \url{https://www.aclweb.org/anthology/D19-3004}
\BIBentrySTDinterwordspacing

\bibitem[Peters et~al.(2018)Peters, Neumann, Iyyer, Gardner, Clark, Lee, and
  Zettlemoyer]{Peters:2018}
\BIBentryALTinterwordspacing
M.~Peters, M.~Neumann, M.~Iyyer, M.~Gardner, C.~Clark, K.~Lee, and
  L.~Zettlemoyer, ``{Deep Contextualized Word Representations},'' in
  \emph{Proceedings of the 2018 Conference of the North American Chapter of the
  Association for Computational Linguistics: Human Language Technologies,
  Volume 1 (Long Papers)}.\hskip 1em plus 0.5em minus 0.4em\relax New Orleans,
  Louisiana: Association for Computational Linguistics, jun 2018, pp.
  2227--2237. [Online]. Available:
  \url{https://www.aclweb.org/anthology/N18-1202}
\BIBentrySTDinterwordspacing

\bibitem[Devlin et~al.(2018)Devlin, Chang, Lee, and Toutanova]{Devlin:2018}
J.~Devlin, M.-W. Chang, K.~Lee, and K.~Toutanova, ``{BERT: Pre-training of Deep
  Bidirectional Transformers for Language Understanding},'' \emph{arXiv
  preprint arXiv:1810.04805}, 2018.

\bibitem[Vaswani et~al.(2017)Vaswani, Shazeer, Parmar, Uszkoreit, Jones, Gomez,
  Kaiser, and Polosukhin]{Vaswani2017}
\BIBentryALTinterwordspacing
A.~Vaswani, N.~Shazeer, N.~Parmar, J.~Uszkoreit, L.~Jones, A.~N. Gomez,
  L.~Kaiser, and I.~Polosukhin, ``{Attention Is All You Need},'' no. Nips,
  2017. [Online]. Available: \url{http://arxiv.org/abs/1706.03762}
\BIBentrySTDinterwordspacing

\bibitem[Cer et~al.(2018)Cer, Yang, Kong, Hua, Limtiaco, John, Constant,
  Guajardo-Cespedes, Yuan, Tar, and Others]{cer2018universal}
\BIBentryALTinterwordspacing
D.~Cer, Y.~Yang, S.-y. Kong, N.~Hua, N.~Limtiaco, R.~S. John, N.~Constant,
  M.~Guajardo-Cespedes, S.~Yuan, C.~Tar, and Others, ``{Universal sentence
  encoder},'' \emph{arXiv preprint arXiv:1803.11175}, 2018. [Online].
  Available: \url{https://arxiv.org/pdf/1803.11175.pdf}
\BIBentrySTDinterwordspacing

\bibitem[Radford et~al.(2018)Radford, Narasimhan, Salimans, and
  Sutskever]{radford2018improving}
A.~Radford, K.~Narasimhan, T.~Salimans, and I.~Sutskever, ``{Improving language
  understanding by generative pre-training},'' \emph{Technical report, OpenAI},
  2018.

\bibitem[Radford et~al.(2019)Radford, Wu, Child, Luan, Amodei, and
  Sutskever]{radford2019language}
A.~Radford, J.~Wu, R.~Child, D.~Luan, D.~Amodei, and I.~Sutskever, ``{Language
  models are unsupervised multitask learners},'' \emph{OpenAI Blog}, vol.~1,
  no.~8, p.~9, 2019.

\bibitem[Liu et~al.(2019)Liu, Ott, Goyal, Du, Joshi, Chen, Levy, Lewis,
  Zettlemoyer, and Stoyanov]{liu:2019-roberta}
\BIBentryALTinterwordspacing
Y.~Liu, M.~Ott, N.~Goyal, J.~Du, M.~Joshi, D.~Chen, O.~Levy, M.~Lewis,
  L.~Zettlemoyer, and V.~Stoyanov, ``{RoBERTa: A robustly optimized BERT
  pretraining approach},'' \emph{arXiv preprint arXiv:1907.11692}, 2019.
  [Online]. Available: \url{https://arxiv.org/pdf/1907.11692.pdf}
\BIBentrySTDinterwordspacing

\bibitem[Yin et~al.(2018)Yin, Deng, Chen, Vasilescu, and Neubig]{yin2018mining}
P.~Yin, B.~Deng, E.~Chen, B.~Vasilescu, and G.~Neubig, ``{Learning to Mine
  Aligned Code and Natural Language Pairs from Stack Overflow},'' in
  \emph{International Conference on Mining Software Repositories}, ser.
  MSR.\hskip 1em plus 0.5em minus 0.4em\relax ACM, 2018, pp. 476--486.

\bibitem[Yao et~al.(2018)Yao, Weld, Chen, and Sun]{yao2018staqc}
Z.~Yao, D.~S. Weld, W.-P. Chen, and H.~Sun, ``{StaQC: A Systematically Mined
  Question-Code Dataset from Stack Overflow},'' in \emph{Proceedings of the
  2018 World Wide Web Conference on World Wide Web}.\hskip 1em plus 0.5em minus
  0.4em\relax International World Wide Web Conferences Steering Committee,
  2018, pp. 1693--1703.

\bibitem[Peddamail et~al.(2018)Peddamail, Wang, Yao, and
  Sun]{peddamail2018comprehensive}
J.~Peddamail, Z.~Wang, Z.~Yao, and H.~Sun, ``{A Comprehensive Study of StaQC
  for Deep Code Summarization},'' in \emph{Proceedings of the 24th ACM SIGKDD
  International Conference on Knowledge Discovery and Data Mining, Lond, UK},
  2018.

\bibitem[Bojanowski et~al.(2016)Bojanowski, Grave, Joulin, and
  Mikolov]{Bojanowski2016}
P.~Bojanowski, E.~Grave, A.~Joulin, and T.~Mikolov, ``{Enriching Word Vectors
  with Subword Information},'' \emph{CoRR abs/1607.04606}, 2016.

\bibitem[Kiros et~al.(2015)Kiros, Zhu, Salakhutdinov, Zemel, Urtasun, Torralba,
  and Fidler]{kiros2015skip}
R.~Kiros, Y.~Zhu, R.~R. Salakhutdinov, R.~Zemel, R.~Urtasun, A.~Torralba, and
  S.~Fidler, ``{Skip-thought Vectors},'' in \emph{Advances in Neural
  Information Processing Systems}, 2015, pp. 3294--3302.

\bibitem[Henderson et~al.(2017)Henderson, Al-Rfou, Strope, Sung, Luk{\'{a}}cs,
  Guo, Kumar, Miklos, and Kurzweil]{henderson2017efficient}
M.~Henderson, R.~Al-Rfou, B.~Strope, Y.-H. Sung, L.~Luk{\'{a}}cs, R.~Guo,
  S.~Kumar, B.~Miklos, and R.~Kurzweil, ``{Efficient natural language response
  suggestion for smart reply},'' \emph{arXiv preprint arXiv:1705.00652}, 2017.

\bibitem[Bowman et~al.(2015)Bowman, Angeli, {Potts Christopher}, and
  Manning]{snli:emnlp2015}
S.~R. Bowman, G.~Angeli, {Potts Christopher}, and C.~D. Manning, ``{A large
  annotated corpus for learning natural language inference},'' in
  \emph{Proceedings of the 2015 Conference on Empirical Methods in Natural
  Language Processing (EMNLP)}.\hskip 1em plus 0.5em minus 0.4em\relax
  Association for Computational Linguistics, 2015.

\bibitem[Vuli{\'{c}} and Moens(2016)]{Vuli2016}
I.~Vuli{\'{c}} and M.-F. Moens, ``{Bilingual Word Embeddings from Comparable
  Data with Application to Bilingual Lexicon Extraction and Word Translation
  Disambiguation},'' \emph{Journal of Artifical Intelligence Research}, 2016.

\bibitem[Johnson et~al.(2017)Johnson, Douze, and J{\'{e}}gou]{JDH17}
J.~Johnson, M.~Douze, and H.~J{\'{e}}gou, ``{Billion-scale similarity search
  with GPUs},'' \emph{arXiv preprint arXiv:1702.08734}, 2017.

\bibitem[Abadi et~al.(2015)Abadi, Barham, Chen, Chen, Davis, Dean, Devin,
  Ghemawat, Irving, Isard, Kudlur, Levenberg, Monga, Moore, Murray, Steiner,
  Tucker, Vasudevan, Warden, Wicke, Yu, and Zheng]{tensorflow2015-whitepaper}
\BIBentryALTinterwordspacing
M.~Abadi, P.~Barham, J.~Chen, Z.~Chen, A.~Davis, J.~Dean, M.~Devin,
  S.~Ghemawat, G.~Irving, M.~Isard, M.~Kudlur, J.~Levenberg, R.~Monga,
  S.~Moore, D.~G. Murray, B.~Steiner, P.~Tucker, V.~Vasudevan, P.~Warden,
  M.~Wicke, Y.~Yu, and X.~Zheng, ``{TensorFlow: Large-Scale Machine Learning on
  Heterogeneous Systems},'' 2015. [Online]. Available:
  \url{http://tensorflow.org/}
\BIBentrySTDinterwordspacing

\bibitem[Kingma and Welling(2014)]{Kingma2014}
D.~Kingma and M.~Welling, ``{Auto-Encoding Variational Bayes},'' in \emph{In
  Proceedings of the International Conference on Learning Representations
  (ICLR)}, 2014, pp. 1--14.

\bibitem[Reimers and Gurevych(2019)]{reimers-gurevych-2019-sentence}
\BIBentryALTinterwordspacing
N.~Reimers and I.~Gurevych, ``{Sentence-BERT: Sentence Embeddings using Siamese
  BERT-Networks},'' in \emph{Proceedings of the 2019 Conference on Empirical
  Methods in Natural Language Processing and the 9th International Joint
  Conference on Natural Language Processing (EMNLP-IJCNLP)}.\hskip 1em plus
  0.5em minus 0.4em\relax Hong Kong, China: Association for Computational
  Linguistics, 2019, pp. 3982--3992. [Online]. Available:
  \url{https://www.aclweb.org/anthology/D19-1410}
\BIBentrySTDinterwordspacing

\bibitem[Ethayarajh(2019)]{ethayarajh2019contextual}
K.~Ethayarajh, ``{How Contextual are Contextualized Word Representations?
  Comparing the Geometry of BERT, ELMo, and GPT-2 Embeddings},'' in
  \emph{Proceedings of the 2019 Conference on Empirical Methods in Natural
  Language Processing and the 9th International Joint Conference on Natural
  Language Processing (EMNLP-IJCNLP)}, 2019, pp. 55--65.

\bibitem[Harris(1954)]{Harris:1954}
Z.~S. Harris, ``{Distributional structure},'' \emph{Word}, vol.~10, no.~23, pp.
  146--162, 1954.

\end{thebibliography}



\end{document}